\documentclass[twocolumn,aps,showpacs,showkeys,amsmath,amssymb]{revtex4}
\usepackage{graphicx}
\usepackage{dcolumn}
\usepackage{bm}
\usepackage{epsf}
\newcommand{\nwl}{\\[2mm]}
\newcommand{\edc}{\end{document}}
\newcommand{\bb} {\begin{thebibliography}}
\newcommand{\eb}{\end{thebibliography}}
\newcommand{\bi}[1]{\bibitem{#1}}
\newcommand{\bc}{\begin{center}}
\newcommand{\ec}{\end{center}}

\newcommand{\be}{\begin{equation}\small}
\newcommand{\ee}{\end{equation}\normalsize}
\newcommand{\bea}{\begin{eqnarray}}
\newcommand{\eea}{\end{eqnarray}}
\newcommand{\ba}{\begin{array}{l}   }
\newcommand{\lab}[1]{\label{#1}}
\newcommand{\ea}{\end{array}}

\newcommand{\dsfrac}{\displaystyle\frac}
\newcommand{\ds} {\displaystyle}

\newcommand{\re}[1]{(\ref{#1})}

\newcommand{\ci}{\cite}

\newcommand{\dsint}{\ds\int}

\newcommand{\vecr}{ \vec{r}}

\newcommand{\half}{\frac{1}{2}}

\newcommand{{\vergul}}{  ,}

\newcommand{\rhozero}{\ds{\rho_0}}
\newcommand{\rhoone}{\ds{\rho_1}}
\newcommand{\sfree}{S_{\mbox{free}}}
\newcommand{\sint}{S_{\mbox{int}}}
\newcommand{\sclas}{S_{\mbox{clas}}}
\newcommand{\veck}{\vec{k}}
\newcommand{\Ek}{E_k}
\newcommand{\vecnab}{\mbox{\boldmath $\nabla$}}
\newcommand{\cale}{{\cal E}}
\newcommand{\calf}{{\cal F}}
\newcommand{\xbar}{{\bar{X}_1}}
\newcommand{\nbar}{{\bar{n}}}
\newcommand{\Sigmabar}{{\ds{\bar{\Sigma}}}}
\newcommand{\mubar}{{\bar{\mu}}}
\newcommand{\vbar}{{\bar{v}}}
\begin{document}
\title{Stability of the homogeneous Bose-Einstein condensate at large gas parameter}
\author{Abdulla Rakhimov$^{a,b}$} \email{rakhimovabd@yandex.ru}
\author{ Chul  Koo Kim$^{a}$} \email{ckkim@phya.yonsei.ac.kr }
\author{ Sang-Hoon  Kim$^{c}$} \email{shkim@mmu.ac.kr }
\author{Jae Hyung Yee$^{a}$} \email{jhyee@phya.yonsei.ac.kr}
\affiliation{$^a$ Institute of Physics and Applied Physics, Yonsei University, Seoul 120-749, R.O. Korea\\
$^b$Institute of Nuclear Physics, Tashkent 702132, Uzbekistan \\
$^c$Division of Liberal Arts and Sciences, Mokpo National Maritime University, Mokpo 530-729, R.O. Korea }

\begin{abstract}
 The properties of the uniform Bose gas is studied within the  optimized variational perturbation theory
 (Gaussian approximation)  in a self-consistent way. It is shown that the  atomic BEC with a
 repulsive interaction becomes unstable when the gas parameter $\gamma=\rho a^3$
exceeds a critical value $\gamma_{crit}\approx 0.01$.
The quantum corrections
beyond the Bogoliubov-Popov  approximation to the energy density, chemical potential and pressure in powers of $\sqrt\gamma$ expansions
 are presented.
\end{abstract}
\maketitle

\section{Introduction}

The long wait after it's prediction (more than 70 years)
for realization of the Bose-Einstein condensate (BEC) is possibly
related to the meta-stability of the initial Bose gas.
In fact, we  require  first,  an atomic system  would stay gaseous
and meta-stable at very low temperature  all the way to the BEC transition,
and  secondly,  development of cooling and trapping techniques to reach
the required regimes of temperature and density \ci{obzorkettel}.
Clearly,  without a proper cooling technique, any ordinary atomic gas would undergo
 into a liquid or a solid state at low temperatures, so a meta-stale state could be created
only with low pressure and weak interaction between atoms.

Even once  created, the condensate still remains as a fragile and subtle object \ci{nozarezkukkitob}.
The enemies of BEC such as crystallization, disassociation and three-body recombination
 may easily destroy it within very short time.
When the sign of interaction (or equivalently of the s-wave scattering length, $a$ )
is suddenly changed into a negative value, the BEC collapses and then undergoes an explosion in which
a substantial fraction of the atoms were blown off ({\it Bosenova}) \ci{donleynature,blochcondmat}.

Due to the  "bad collisions", even an atomic BEC  with a repulsive interaction has a limited life time.
Recently, Cornish {\it et. al.} \ci{cornish} carried out an ingenious
 experiment with spin
polarized atomic $^{85}$Rb. In the experiment, they showed that one could control
 the strength of interatomic interaction for the BEC by employing the  Feshbach resonance method.
 A very large value of the scattering length $(a \approx 4500\AA)$ has been achieved
in this experiment, which corresponds to  the gas parameter of the condensate to be about
$\gamma_{max}\approx 0.01$.
	This phenomenon has been recently studied by Yin \ci{Yin}
in Random Phase Approximation.
 The author has shown that when $\gamma$ exceeds a certain critical value the molecular
excitation energy becomes  imaginary and, hence, the atomic BEC is dynamically unstable
against molecular formation.

It is well known that many of the basic properties of the condensate of dilute Bose gases
in existing experiments can be described reasonably well using the mean field approximation
(MFA) which reduces the problem to the  classical Gross - Pitaevskii equation (GPE) \ci{pitbook}.
However, fluctuations of the quantum field around the mean field provide corrections which
become increasingly important as higher condensate densities (say large $\gamma$ ) are achieved.
It is therefore important to understand the effects of quantum field fluctuations especially
at large gas parameters.

In the present paper,  we study the properties of a homogeneous atomic Bose gas
using  optimized Gaussian approximation \ci{gauss}.
It has been proven that  the corresponding  Gaussian effective potential contains one loop,
sum of all daisy and superdaisy graphs of perturbation theory \ci{amelino}
and  leading order in  $1/N$ expansion.

The first application of the Gaussian variational approach to a uniform BEC was done
by Bijlsma and Stoof ten years ago \ci{blismastoof}. However,  it was pointed out
in excellent review by Andersen that  \ci{andersen} even a modified (by introducing many body T-matrix)
Gaussian approximation of Ref. \ci{blismastoof} does not satisfy the Hugenholtz-Pines (H-P) theorem  especially at very low temperatures.
 This is particularly caused by a long standing problem encountered in the  most of
field theoretical approximations: it is impossible to satisfy the H-P theorem, namely   making the theory gapless
at the same time and  maintaining the number of particles  with the same value of the chemical potential.
 In other words, the chemical potential defined by the H-P theorem
does not coincide with the chemical potential found from the minimization
of the thermodynamic potential with respect to the condensate density.
Note that,  even the T-matrix approximation cannot resolve this problem completely since in this case
one gets ``mismatch of approximations" which makes the approach as non
self-consistent.

One of the possible solutions of the above mentioned problem has been proposed recently
by Yukalov \ci{yukalov}. He has shown that  Hartree-Fock approximation (HFA)
can be made both conserving and gapless by taking into account of two normalization conditions
instead of one. Hence,  two chemical potentials each for the condensed fraction ($\mu_0$)
and the uncondensed fractions ($\mu_1$) should be introduced to describe the BEC
self-consistently.

In the present paper,  we reformulate the field theoretical Gaussian approximation
following the Yukalov prescription, and apply this self-consistent approach to investigate
the properties of a uniform BEC.

The paper is organized as follows:
In Sect. II, we extend the field theoretical approach by implementing  Yukalov's ideas.
In Sect. III, we calculate the free energy in a Gaussian approximation, and also show
it's relation to the one loop and Bogoliubov-Popov approximations (BPA).
In the next two sections, we present  the procedure of minimization of the free energy.
The numerical results and their discussion are  presented in section VI.
Sect. VII  summarizes the paper.

\section{Quantum Field Formulation with Yukalov prescription}

A grand canonical ensemble of Bose particles with a short range s - wave interaction
is governed by the Euclidian action \ci{andersen},
\footnote
{In the remainder of the paper, we set
 $\hbar=1$ and $k_B=1$ for convenience.
}
\begin{eqnarray}
S[\psi,\psi^{*}]=\dsint_{0}^{\beta}d\tau\dsint d\vecr
{\Large\{}
\psi^{*}(\tau,r)[\partial_\tau-\dsfrac{\vecnab^2}{2m}-\mu]\psi(\tau,r)
\nonumber \\
 + \dsfrac{g}{2}[\psi^{*}(\tau,r)\psi(\tau,r)]^2
{\Large\}},
\lab{eq1}
\end{eqnarray}
where $\psi^{*}(\tau,r)$ is a complex field operator  that creates  a boson
at the position $\vecr$, $\mu$  the chemical potential,
$g$  the  coupling constant given by $4\pi a/m$,
 $m$  the atomic mass
 and  $\beta=1/T$  the inverse of temperature $T$.
The free energy  of the system can be determined as
\be
\calf(\mu)=- T\ln Z,
 \lab{eq2}
\ee
  where $Z$ is the functional integral,
\be
Z=\ds{\dsint \cal{D} \psi \cal{D} \psi^{*} }\exp\{-S[\psi,\psi^{*}] \},
 \lab{eq3}
\ee
 performed all over Bose fields $\psi$ and $\psi^{*}$ periodic in $\tau\in[0,\beta]$.
When the temperature in a Bose system falls  below the condensation temperature $T_c$,
breaking of the $U(1)$ gauge symmetry may be taken into account by the Bogoluibov  shift of the field operator,
 \be
 \psi(\tau,r)=v(\tau,r)+\tilde{\psi}(\tau,r),
 \lab{eq4}
 \ee
 where $v(\tau,r)$ is  the condensate order parameter. In  the uniform system
  $v(\tau,r)$ is a real constant, $v(\tau,r)=v,$
 $\tilde{\psi}(\tau,r)$ is the field operator of the uncondensed
 particles satisfying the same Bose  commutation relation as
 ${\psi}(\tau,r)$. The conservation of particle numbers
 requires that $\tilde{\psi}(\tau,r)$  has non -  zero momentum component so that
 \be
 \langle \tilde{\psi} \rangle =0,
 \lab{eq5}
 \ee
 and $\tilde{\psi}$ and $v$ are orthogonal each other
 \be
 \dsint d \vecr \tilde{\psi}(r)v(r)=0.
 \lab{eq6}
 \ee
 The condensate order parameter $v$ defines the density of  condensed particles
while $\tilde{\psi}$ defines the density of  uncondensed particles:
 \be
 \rhozero=v^2, \quad
\rhoone=< {\tilde{\psi}}^{*}(r) \tilde{\psi}(r)>.
\lab{eq7}
 \ee

Having performed the Bogoliubov shift, one may introduce the  grand canonical thermodynamic potential
of the system $\Omega$ as
\be
\Omega(\mu,v)=\ds{\calf (\mu,v)|_{v=\langle \psi \rangle}}.
\lab{aeq9}
\ee
In a stable equilibrium, $\Omega$ attains the minimum:
\be
\dsfrac{d\Omega(\mu,v)}{dv}=0, \quad
\dsfrac{d^2 \Omega(\mu,v) }{d^2 v} > 0.
\lab{eq8}
\ee

Apart from the H-P theorem, the chemical potential should satisfy
 the normalization condition
\be
N= \langle \dsint d \vecr {\psi}^{*}(r) {\psi}(r) \rangle,
 \lab{eq10}
\ee
where $N$ is the total number of particles. However, as it was pointed out in the
above, the chemical potential
corresponding to the minimum of $\Omega$ may not  correspond
to the chemical potential $\mu$ determined from the normalization condition.

To overcome this difficulties,  Yukalov \ci{yukalov} proposed to
\begin{itemize}
\item
Introduce one more normalization condition $N_0=\rhozero V$.  So that for the uniform system
 \be
N_0+N_1=N, \quad
N_1= \langle \dsint d\vecr {\tilde{\psi}}^{*}(r) \tilde{\psi}(r) \rangle,
 \lab{eq11}
 \ee
which simply states that the total number of particles should be  equal to
the sum of the number of condensed and uncondensed particles.
\item
 Introduce two chemical potentials $\mu_0$ and $\mu_1$, for the condensed and the uncondensed
fractions respectively as well as a Lagrange multiplier $\Lambda$ to satisfy the Eq. \re{eq5}.
The total system chemical potential,
$\mu=-(\partial \Omega/ \partial N)$,
is given by
\be
\mu=\dsfrac{\mu_0 N_0+\mu_1 N_1}{N}.
 \lab{eq12}
\ee
\end{itemize}

These prescriptions lead to the following action,
\begin{eqnarray}
S[\psi,\psi^{*}]=\dsint_{0}^{\beta}d\tau\dsint d\vecr
{\Large \{}
\psi^{*}(\tau,r)[\partial_\tau-\dsfrac{\vecnab^2}{2m}]\psi(\tau,r)
\nonumber \\
- \mu_1{\tilde{\psi}}^{*}(\tau,r) \tilde{\psi}(\tau,r)-\mu_0v^2
-\Lambda\tilde{\psi}(\tau,r)
\nonumber \\
- \Lambda^{*}{\tilde{\psi}}^{*}
(\tau,r)+\dsfrac{g}{2}[\psi^{*}(\tau,r)\psi(\tau,r)]^2
 {\Large \}},
\lab{eq13}
\end{eqnarray}
 which should be  used in Eq. \re{eq3}.
Further, $\mu_0$ can be determined from the minimum condition  Eq. \re{eq8},
while  $\mu_1$ itself by the requirement of H-P theorem
\be
\mu_1=\Sigma_{11}-\Sigma_{12},
 \lab{eq15}
 \ee
  where $\Sigma_{11}$ and $\Sigma_{12}$ are the normal and the anomalous
  self-energies.
  As to the condensed fraction $N_0$,
  it could be found by solving the normalization Eq. \re{eq11} where
the uncondensed fraction $N_1$ is given by
\be
N_1=-\left(\dsfrac{\partial \Omega}{\partial \mu_1}\right).
\lab{eq14}
\ee

\section{Gaussian, one - loop and Bogoliubov-Popov approximations }

In the present section, we show how this scheme can  be realized in
practice.
Substituting Eq. \re{eq4} into Eq. \re{eq13},
 one may rewrite the action in powers of $v$ and ${\tilde{\psi}},$
\begin{eqnarray}
S &=& S^{(0)}+ S^{(1)}+S^{(2)}+S^{(3)}+S^{(4)}.
\nonumber \\
S^{(0)}&=& \dsint_{0}^{\beta}d\tau\dsint d\vecr
  \{-v^2 \mu_0+\dsfrac{gv^4}{2} \},
\nonumber \\
S^{(1)}&=& \dsint_{0}^{\beta}d\tau\dsint d\vecr
   \,[gv^3-\Lambda^*-\Lambda][{\tilde{\psi}}^{*}+{\tilde{\psi}} ],
\nonumber \\
S^{(2)}&=& \dsint_{0}^{\beta}d\tau\dsint d\vecr \{
{\tilde{\psi}}^{*}[\partial_\tau -\dsfrac{\vecnab^2}{2m}-\mu_1]{\tilde{\psi}}
\nonumber \\
& & +\dsfrac{gv^2}{2}[{\tilde{\psi}}^{*}{\tilde{\psi}}^{*}+
4{\tilde{\psi}}^{*}{\tilde{\psi}}+{\tilde{\psi}}{\tilde{\psi}}]
\},
\nonumber \\
 S^{(3)}&=& g\dsint_{0}^{\beta}d\tau\dsint d\vecr
v\{{\tilde{\psi}}^{*}{\tilde{\psi}}^{*}{\tilde{\psi}}
+{\tilde{\psi}}^{*}{\tilde{\psi}}{\tilde{\psi}} \},
\nonumber \\
S^{(4)}&=& \dsfrac{g}{2}\dsint_{0}^{\beta}d\tau\dsint d\vecr
{\tilde{\psi}}^{*}{\tilde{\psi}}^{*}{\tilde{\psi}} {\tilde{\psi}}.
 \lab{eq16}
\end{eqnarray}
In the following,  $S^{(1)}$ will be omitted since it can be
set to zero by an appropriate choice of $\Lambda$ in order to satisfy
Eq. \re{eq5}.

Now in accordance with the  variational perturbation theory, we add and
subtract the following term:
 \be
S^{(\Sigma)}=\dsint_{0}^{\beta}d\tau\dsint d\vecr \left[
\Sigma_{11}{\tilde{\psi}}^{*}{\tilde{\psi}}+\dsfrac{1}{2}\Sigma_{12}
({\tilde{\psi}}^{*}{\tilde{\psi}}^{*}
+{\tilde{\psi}}{\tilde{\psi}}) \right],
 \lab{eq17} \ee
 assuming $\Sigma_{11}$ and $\Sigma_{12}$ as
real constants.
Further, we write  the quantum fluctuating
field ${\tilde{\psi}}$ in terms of two real fields
 \be
{\tilde{\psi}}=\dsfrac{1}{\sqrt{2}}(\psi_1+i\psi_2),\quad
{\tilde{\psi}}^{*}=\dsfrac{1}{\sqrt{2}}(\psi_1-i\psi_2).
\lab{eq18}\ee
After some algebraic manipulations \ci{nagaosa,haugset},
 one can split the action into ``classical",
``free", and ``interaction" parts:
\begin{eqnarray}
 S&=&\sclas+\sfree+\sint.
\nonumber \\
\sclas&=&V\beta  ( -v^2 \mu_0+\dsfrac{gv^4}{2} ),
\nonumber \\
\sfree&=&\dsfrac{1}{2}\dsint_{0}^{\beta}d\tau\dsint d\vecr [
i\epsilon_{ab}\psi_a\partial_\tau \psi_b
+\psi_1(-\dsfrac{\vecnab^2}{2m}+X_1)\psi_1
\nonumber \\
& & +\psi_2(-\dsfrac{\vecnab^2}{2m}+X_2 ) \psi_2 ].
 \\
\sint&=&\sint^{(2)}+\sint^{(3)}+\sint^{(4)}.
\nonumber \\
\sint^{(2)}&=&\dsfrac{1}{2}\dsint_{0}^{\beta}d\tau\dsint d\vecr [
\psi_{1}^{2}(3gv^2-\Pi_{11})+\psi_{2}^{2}(gv^2-\Pi_{22}) ],
\nonumber \\
\sint^{(3)}&=&\dsfrac{g}{\sqrt{2}}\dsint_{0}^{\beta}d\tau\dsint
d\vecr  v\psi_{1}(\psi_{1}^{2}+\psi_{2}^{2}),
\nonumber \\
\sint^{(4)}&=&\dsfrac{g}{8}\dsint_{0}^{\beta}d\tau\dsint d\vecr
(\psi_{1}^{4}+2\psi_{1}^{2}\psi_{2}^{2}+\psi_{2}^{4}).
 \lab{eq19}
\end{eqnarray}

 Here,  $\epsilon_{ab} (a,b=1,2)$ is the antisymmetric tensor in two
 dimensions with $\epsilon_{12}=1$ and  following notations
 are introduced,
 \be  \ba
 \Pi_{11}=\Sigma_{11}+\Sigma_{12},\quad\quad
 \Pi_{22}=\Sigma_{11}-\Sigma_{12},\nwl
 X_1=\Pi_{11}-\mu_{1}, \quad\quad X_2=\Pi_{22}-\mu_{1}.
\lab{eq20}
 \ea  \ee
In accordance with Refs. \ci{andersen,braatentrap}, $\Pi_{ab}$
are the components of the  $2\times 2 $ self-energy matrix.

The free part of the action,   $\sfree$ in Eq. \re{eq19} gives rise to
a propagator, which can be used in perturbative framework.
In a momentum space,
 \be \tilde{\psi}_a(\tau,r)=\dsfrac{1}{\sqrt{ \beta V
}}\sum_{n=-\infty}^{\infty}\sum_{k}\tilde{\psi}_a(\omega_n,\veck)
\exp\{i\omega_n\tau+i\veck\vecr\},
 \lab{21}\ee
  where
$\ds{\sum_{k}}=V\int d\veck/(2\pi)^3$, and
$\omega_n=2\pi nT$ is the  Matsubara frequency.
The propagator  is given by
\begin{eqnarray}
G(\omega_n,k)=\dsfrac{1}{\omega_{n}^{2}+\Ek^2}\left(
\begin{array}{cc}
\varepsilon_k+X_2& \omega_n\\
\nonumber
 -\omega_n&\varepsilon_k+X_1
\end{array}\right),\\
 \\ \nonumber
\label{eq22}
\end{eqnarray}
with the dispersion relation,
$\Ek^2=(\varepsilon_k+X_1)(\varepsilon_k+X_2)$ and
$\varepsilon_k=\ds{\vec{k}}^2/2m$.

With this Green's function using Eqs. \re{eq2} and \re{eq3}, and neglecting terms $\sint^{(3)}$ and $\sint^{(4)}$,
 one may get the thermodynamic potential
 in the one loop approximation:
\be \ba
{\Omega^{(1L)}(\mu_0,\mu_1,v)}= V \left(-\mu_0v^2+\dsfrac{gv^4}{2}\right)
\nwl
+\dsfrac{1}{2}\ds{\sum_{k}}\Ek+
 T\ds{\sum_{k}}\ln[1-e^{-\beta\Ek}]
\nwl
+ \dsfrac{1}{2}\left[ B(3gv^2-\Pi_{11})+A(gv^2-\Pi_{22}) \right]
\lab{eq23},
\ea\ee
 with $\Pi_{11}=3gv^2$ and $\Pi_{22}=gv^2$ (
$A$ and $B$ will be given below), so that the last term in square
bracket can be  dropped.
Note that, hereafter we perform explicit summation by Matsubara frequencies
(see e.g. \ci{haugset}).
As to the BPA, it can be obtained
by introducing an auxiliary expansion parameter
$\eta_{1L}$ as it was shown by Kleinert \ci{klcondmat}.

Loop expansion of $\Omega$ may be organized by using the propagator
$G(\omega_n,\veck)$ with constraints $X_1=2gv^2$ and $X_2=0$ as illustrated
in Ref. \ci{braatenepj}.
To take into account higher order  quantum fluctuations, one has to calculate
$\langle \sint^{(3)}\rangle$ and $\langle \sint^{(4)}\rangle$.
Although these quantities can not be evaluated exactly,
they may be estimated in the Gaussian approximation
\footnote{The details of the calculations will be given in a
separate paper.}, where for the homogeneous system:
  \be \ba
\langle
\sint^{(3)} \rangle=0,
\nwl \langle \psi_{a}^{2}
\rangle=G_{aa}(r-r')|_{r\rightarrow r'}\equiv G_{aa}(0),
\nwl
\langle \psi_{1}^{2}\psi_{2}^{2} \rangle=\langle \psi_{1}^{2}
\rangle\langle \psi_{2}^{2} \rangle,\quad
\langle
\psi_{a}^{4} \rangle=3G_{aa}^{2}(0),
\nwl
G_{11}(0)=\dsfrac{1}{V\beta}
\sum_{n=-\infty}^{\infty}\sum_{k}G_{11}(\omega_n,\veck)=
V^{-1}B,
\nwl G_{22}(0)=V^{-1}A.
\lab{eq24}
 \ea\ee
Finally, combining Eqs. \re{eq23} and \re{eq24},
 we get the following  expressions for the thermodynamic potential:
\be \ba
{\Omega (X_1,X_2,v,\mu_0,\mu_1)}=V(-\mu_0v^2+\dsfrac{gv^4}{2})
\nwl
+\dsfrac{1}{2}\ds{\sum_{k}}\Ek
+T\ds{\sum_{k}}\ln[1-e^{-\beta\Ek}]
\nwl
+ \dsfrac{1}{2} \left[ B(3gv^2-\Pi_{11})
+A(gv^2-\Pi_{22}) \right]
\nwl
+\dsfrac{g\rho}{8N}\left[ 3(A^{2}+B^{2})+2AB \right],
\lab{eq25}
\ea\ee
where
\be \ba
A\equiv\ds\sum_{k}\dsfrac{\varepsilon_k+X_1}{\Ek}\ds{
\left[ \dsfrac{1}{2} +\dsfrac{1}{\exp(\beta\Ek)-1} \right] },
\nwl\nwl
B\equiv\ds\sum_{k}\dsfrac{\varepsilon_k+X_2}{\Ek}\ds{ \left[ \dsfrac{1}{2}
+\dsfrac{1}{\exp(\beta\Ek)-1}
\right]}.
 \lab{eq26}
\ea \ee

The free energy in Eq. \re{eq25} is supposed to have all the
information about the system. Particularly taking it's
derivative with respect to $\mu_1$,
 one gets the expression
for the uncondensed fraction $N_1$:
 \be \ba
N_1=-\left(\dsfrac{\partial \Omega}{\partial \mu_1}\right)
\nwl
=\dsfrac{1}{2}\{
A+B-(3gv^2-\Pi_{11})B'
-(gv^2-\Pi_{22})A'
\nwl
-\dsfrac{g}{2V}[(3A+B)A'+(3B+A)B'] \}
\lab{eq27},
\ea\ee
where $A'={\partial A}/{\partial \mu_1}$ and
$B'={\partial B}/{\partial \mu_1}$.
Note that the same expression for the  uncondensed fraction could be obtained
 in an alternative way as
 \be
 \ba
 \rho_1=\dsfrac{N_1}{V}=<\tilde\psi^{*}\tilde\psi>=\dsfrac{1}{Z}{\ds{\dsint \cal{D}
 \tilde\psi \cal{D} \tilde\psi^{*} }\exp\{-
S[\psi,\psi^{*}] \}\tilde\psi^{*}\tilde\psi}
\ea
\ee
\section{The gap equations  and the thermodynamic potential at $T=0$}

In this section, the variational parameters $\Pi_{11}$ and $\Pi_{22}$
will be determined using the  principle of minimal sensitivity \ci{gauss}.
 From Eqs. \re{eq25} and \re{eq26},
the gap equations may be found
\be \ba
\dsfrac{\partial\Omega (X_1,X_2,v,\mu_0,\mu_1)}{\partial X_1}
\nwl =\dsfrac{1}{2}\{
A'_1[gv^2-\mu_1-X_2]  + B'_1[3gv^2-\mu_1-X_1]
\nwl + g[A'_1(3A+B)+B'_1(3B+A)]/2V
 \}=0,
  \lab{eq28}
  \ea\ee
 and
 \be \ba
 \dsfrac{\partial {\Omega (X_1,X_2,v,\mu_0,\mu_1)}}{\partial X_2}
 \nwl =\dsfrac{1}{2}\{
A'_2[gv^2-\mu_1-X_2] + B'_2[3gv^2-\mu_1-X_1]
\nwl +
g[A'_2(3A+B)+B'_2(3B+A)]/2V
 \}=0.
 \lab{eqq28}
 \ea\ee
 where
 \be\ba\
A'_1\equiv\dsfrac{\partial A}{\partial
X_1}=\dsfrac{1}{4}\ds\sum_k\dsfrac{1}{\Ek}=\dsfrac{\partial
B}{\partial X_2}\equiv B'_2,
\nwl
 A'_2\equiv\dsfrac{\partial
A}{\partial
X_2}=-\dsfrac{1}{4}\ds\sum_k\dsfrac{(\varepsilon_k+X_1)^2}{\Ek^3},
\nwl
B'_1\equiv\dsfrac{\partial B}{\partial
X_1}=-\dsfrac{1}{4}\ds\sum_k\dsfrac{(\varepsilon_k+X_2)^2}{\Ek^3}.
 \lab{eq29}
 \ea\ee

 Above,   we have two equations \re{eq28} and \re{eqq28}
 with respect to three  unknown quantities
 $\{X_1, X_2, \mu_1\}$. An additional equation
  is supplied from the  relation
 between the chemical potential
 and the self-energies given by the H-P theorem.
  So, from Eqs. \re{eq15} and \re{eq20},
   one can immediately  conclude $X_2=0$,
   and, hence,  in the long wavelength limit
  $(k\rightarrow 0)$,
 the quasiparticle energy $\Ek$ behaves as $ck$ (with $c=\sqrt{X_1/2m}$ )
 thus being gapless, as expected.
With this constraint, the gap equations may be  simplified as
\be
2X_1+\mu_1-5gv^2-\frac{11}{8}\dsfrac{g}{V}I_{1,1}(X_1)=0,
 \lab{eq30}\ee
 \be
 \ba
I_{0,1}(X_1)[3gv^2-\mu_1-X_1]-I_{-2,-1}(X_1)[gv^2-\mu_1]
\nwl + \dsfrac{gI_{1,1}(X_1)}{8V}
[5I_{0,1}(X_1)+I_{-2,-1}(X_1)]=0,
 \lab{eq31}
 \ea
 \ee
and  Eq. \re{eq27} as
\be \ba
N_1=\dsfrac{1}{8}I_{1,1}(X_1)
\nwl +
\dsfrac{[I_{0,1}(X_1)-2I_{-2,-1}(X_1)][8Vgv^2-gI_{1,1}(X_1)-8V\mu_1]
}{128Vm}.
 \lab{eq32} \ea\ee
 Here, the following dimensionless integral
\be
I_{i,j}(X_1)=\ds\sum_k\dsfrac{\varepsilon_{k}^{i}m^{j-i}}{\Ek^j}
\lab{eq33} \ee
is introduced. Their explicit expressions  and the relations between
them
evaluated in dimensional regularization are presented in the Appendix of Ref. \ci{andersen}.
In particular,
\be
\ba
I_{1,1}(X_1)=\frac{V(2mX_1 )^{3/2}}{3\pi^2} =2B|_{X_2=0}=-4A|_{X_2=0},
\nwl
\nwl
\dsfrac{dI_{1,1}(X_1)}{dX_1}=-\dsfrac{I_{0,1}(X_1)}{m}.
\lab{eqI33}
\ea
\ee

Note that Eqs. \re{eq31} and \re{eq32} include $I_{-2,-1}(X_1)$,
which is infrared  divergent,  $I_{-2,-1}(X_1)\sim
1/\epsilon+\ln\kappa^2/mX_1$ (with $\epsilon\rightarrow 0$).
Below
we show that this integral will be canceled exactly. In fact,
eliminating $X_1$ from Eq. \re{eq30} as
 \be
X_1=-\dsfrac{\mu_1}{2}+\dsfrac{5gv^2}{2}+\dsfrac{11gI_{1,1}(X_1)}{16V},
\lab{eq34}\ee
and substituting it into Eq. \re{eq31}, one  observes that the latter
is factorized:
 \be
[I_{0,1}(X_1)-2I_{-2,-1}(X_1)][gI_{1,1}(X_1)-8Vgv^2+8V\mu_1]=0.
\lab{eq35} \ee
 Finally, from the last two equations, we find the  formal
solutions of the gap equations
\be
X_1=2gv^2+\dsfrac{3g}{4V}I_{1,1}(X_1),
\lab{eq36}
\ee
 \be
\mu_1=gv^2-\dsfrac{g}{8 V}I_{1,1}(X_1).
\lab{eq37}\ee
We denote these optimum values of $X_1$ and $\mu_1$ by $\xbar$ and
$\mubar_1$, respectively, which are explicitly dependent on  $v^2$.  Now,
comparing Eqs. \re{eq32} and  \re{eq35},
 one can easily see that only
the first term in Eq. \re{eq32} survives
 \be
N_1=\dsfrac{1}{8}I_{1,1}(X_1).
\lab{eq38}\ee

Now, inserting these  formal solutions into Eq. \re{eq25}
and using the relations between the integrals, Eq. \re{eqI33},
gives  the following form for $\Omega$
 \be \ba
\Omega(\xbar,v,\mu_0)=V(-\mu_0v^2+\dsfrac{gv^4}{2})
+\dsfrac{m}{2}I_{0,-1}(\xbar)
\nwl
- \dsfrac{11gI_{1,1}^{2}(\xbar)}{128V},
\lab{f40}\ea\ee
 where
 \be
 \ba
I_{0,-1}(\xbar)=\dsfrac{1}{m}\ds{\sum_k
\sqrt{\varepsilon_k}\sqrt{\varepsilon_k+\xbar}}
=\dsfrac{2\sqrt{2}V(m\xbar)^{5/2}}{15m^2\pi^2}.
\ea
\ee
In particular, neglecting in Eq. \re{f40} the last term gives
the one-loop result:
\be
\ba
\Omega(\xbar,v,\mu_0)|_{\xbar=2gv^2}=\Omega^{(1L)}(\mu_0,\mu_1,v)|_{\mu_1=gv^2}.
\ea
\ee
presented in the previous section.
In the stable equilibrium,  the
 grand canonical potential reaches the global minimum
 as a function of $v:$
  \be\ba
   \dsfrac{d\Omega(\xbar,v,\mu_0)}{d n_{0}}=V\rho(-\mu_0+g\rho n_0)
   \nwl +\dsfrac{X'_1 I_{1,1}(\xbar)}{4}\ds{
[1+\dsfrac{11gI_{0,1}(X_ 1)}{16Vm} ] }=0
\lab{f41}\ea\ee
 where
$n_0=v^2/\rho=N_0/N$ and  $X'_1=({d \xbar}/{d n_0})$.
Note that  the same equation could be obtained from the original
equation \re{eq25} as:
\be \ba
\dsfrac{d {\Omega (X_1,X_2,v,\mu_0,\mu_1)}}{dn_0}=\dsfrac{\partial\Omega}{\partial n_0}
\nwl +\ds{\left(\dsfrac{\partial\Omega}{\partial \mu_1}\right)\dsfrac{\partial\mu_1}{\partial n_0}}+
 \ds{\left(\dsfrac{\partial\Omega}{\partial X_1}\right)\dsfrac{\partial X_1}{\partial n_0}}+
  \ds{\left(\dsfrac{\partial\Omega}{\partial X_2}\right)\dsfrac{\partial X_2}{\partial n_0}}=0,
 \lab{f40OM}
\ea
\ee
where the last two terms may be omitted due to the gap
Eqs. \re{eq28} and \re{eqq28}, and the factor
in the second term is related to $N_1$ by \re{eq14}.

Clearly, the optimal value of $v^2$, i.e. $\vbar^2$   defined by Eq. \re{f41}, should correspond to the normalization condition in Eq. \re{eq11} (constraint):
\be
\vbar^2+\rho_1(\xbar)=\vbar^2+\dsfrac{I_{1,1}(\xbar)}{8V}=\rho,
\lab{f39}\ee
 which may be considered as a nonlinear equation with respect to
 the c-number
$\vbar^2$ with a fixed $\rho$ and  $\xbar(\vbar^2)$.

 Strictly speaking, $\vbar^2$ must be determined from Eq. \re{f41} as a function
 of $\mu_0$, and after substituting it into Eq. \re{f39},
  the latter should be solved
 with respect to $\mu_0$. However, this would be a rather complicated way,
 since Eq. \re{f41}  is a highly nonlinear equation.
On the other hand, one may assume that
 $\vbar^2$ is known as a solution of Eq. \re{f39}
 and $\mu_0 $ could be extracted from Eq. \re{f41}.

 Following  this strategy, we obtain
  \be \ba \mubar_0=g\rho
\nbar_0+\dsfrac{X'_1 I_{1,1}(\xbar)}{4V\rho}\ds{
\left[1+\dsfrac{11gI_{0,1}(\xbar)}{16Vm} \right] },
 \lab{f42}\ea\ee
in particular, neglecting the second term in square brackets
and taking into account $X_{1}^{1L}=2g\rho n_0$,
we have $\mu_{0}$ for the one-loop approximation
\be
\mu_{0}^{1L}=g\rho \left[ {n_0+\dsfrac{I_{1,1}(\xbar^{1L})}{2V\rho}}  \right].
\lab{f43}\ee
Further simplification, by introducing an auxiliary expansion parameter
$\eta_{1L}$ as in Ref. \ci{klcondmat}, gives   $\mu_0$ for the BPA
\be
\mu_{0}^{BP}=g\rho\left[1-\dsfrac{gI_{0,1}(X_ {1}=2g\rho)}{2Vm}  \right].
\lab{f44}
 \ee

As to the total system chemical potential $\mu$, it follows from Eqs.
\re{eq11} and \re{eq12} as
\be \mu=\mubar_1\nbar_1+\mubar_0\nbar_0,
\lab{f430}
\ee
where $\nbar_1=1-\nbar_0$,
and $\mubar_1$ and $\mubar_0$ are
given by Eqs. \re{eq37} and \re{f42}, respectively.
Now, substituting \re{f42} into \re{f40}, one may obtain the pressure as $P=-\Omega/V$
\be
\ba
P=\half gn_{0}^{2}\rho^2+\dsfrac{1}{4V}[n_0\xbar'I_{1,1}(\xbar)-2mI_{0,-1}(\xbar)]\nwl
\\
+\dsfrac{11gI_{1,1}(\xbar)}{128mV^2}[2n_0\xbar'I_{0,1}(\xbar)+mI_{1,1}(\xbar)].
\lab{davlen}
\ea
\ee

The ground state energy density of the BEC, $\cale$, may be obtained by
a well known formula $\cale=(\Omega+\mu N)/V$. This may be easily
done by rewriting the term $\mu_0v^2 V$ in Eq. \re{f40} as $\mu_0v^2
V=\mu N-\mu_1 n_1N$ (which follows from Eq. \re{f43}) and using
Eq. \re{eq37}:
\be \ba
 \Omega(\xbar,\vbar)=-\mu
N+\dsfrac{V\rho^2g\nbar_{0}^{2}}{2}+\dsfrac{m}{2}I_{0,-1}(\xbar)
\nwl
+\dsfrac{\rho g \nbar_0}{8}I_{1,1}(\xbar)
-\dsfrac{13g}{128V}I_{1,1}^{2}(\xbar).
 \lab{f440}
 \ea\ee
  Now, one may immediately obtain
 \be \ba
\cale=\dsfrac{g\vbar^4}{2}+\dsfrac{m}{2V}I_{0,-1}(\xbar)+\dsfrac{g
\vbar^2}{8V}I_{1,1}(\xbar)-\dsfrac{13g}{128V^2}I_{1,1}^{2}(\xbar),
\lab{f45}
\ea\ee
 and
  \be \ba
{\cale}^{BP}=\dsfrac{g\rho^2}{2}+\dsfrac{m}{2V}\ds{I_{0,-1}(X_
1)|_{X_ 1=2g\rho}}, \lab{f46}\ea\ee for the Gaussian and Bogoliubov - Popov
approximations respectively.

It is well known that in the BPA, the normal
$(\Sigma_{11})$ and the anomalous  self-energies $(\Sigma_{12})$ are
rather simple \ci{klcondmat}:
\be \Sigma_{11}^{BP}=2g\rho ,
\quad\quad \Sigma_{12}^{BP}=g\rho.
\lab{f47}\ee
In the Gaussian approximation using
Eqs. \re{eq20}, \re{eq36}, and \re{eq37}, one  obtains
 \be \ba
\Sigma_{11}=\dsfrac{X_1}{2}+\mu_1=2gv^2+\dsfrac{g}{4V}I_{1,1}(X_
1),
\nwl
\Sigma_{12}=\dsfrac{X_1}{2}=gv^2+\dsfrac{3g}{8V}I_{1,1}(X_
1),
 \lab{f48}\ea\ee
 which can be further simplified  at the stationary point  as
  \be \ba
  \Sigmabar_{11}=2g\rho\nwl
\Sigmabar_{12}=g\rho(1+2\nbar_1).
\lab{f49}
\ea\ee
 Clearly, neglecting the
uncondensed fraction $\nbar_1$ in the last equation, we recover the
 Bogoliubov - Popov approximation, \re{f47}. The dimensionless sound
velocity defined as $c=\ds{\lim_{k\rightarrow
0}\Ek/k=\sqrt{\xbar/2m}}$ is simply related to $\Sigma_{12}$ as \be
c^2=\dsfrac{\Sigmabar_{12}}{m}.
 \lab{f50}\ee

\section{Solutions to the gap equations}

In this section, we analysis possible solutions to the gap
equation \re{eq36} which can be written as
\be
X_1=2gv^2+\dsfrac{g(mX_1)^{(3/2)}}{\sqrt{2}\pi^2}.
 \lab{eq39}\ee
Before  solving this  equation, we emphasize  that in accordance with the general principle of the variational Gaussian approximation, the constraint in Eq.
\re{f39} and the procedure of  minimization of the free energy with respect to $v^2$
may be imposed only after finding an explicit expression for
$X_1\equiv X_1(v^2)$ as a function of $v^2$,
which can be done by  solving Eq. \re{eq39} analytically.
Note that when the second term on the RHS of Eq. \re{eq39} is neglected, one obtains
a well known result of the  one-loop approximation:
$X_{1}^{1L}=2gv^2$,
and further, assuming here $v^2=\rho$ gives the self-energy for
BPA : $X_{1}^{BP}=2\rho g$.


In general, the Eq. \re{eq39} can be rewritten in a dimensionless form
\be
N_{\gamma}=\dsfrac{432Z}{\pi}
-3456\ds{{\left(\dsfrac{Z}{\pi}\right)}^{3/2}},
\lab{eq40}\ee
where the following dimensionless quantities were
introduced
 \be \ba
 Z=\gamma X_1/2g\rho, \quad
N_\gamma=\dsfrac{432\gamma n_0}{\pi},
 \lab{eq41} \ea\ee
with $\gamma=a^3\rho$ is the gas parameter.
 Analysis shows
that Eq. \re{eq40} has no real positive solution when
$N_{\gamma}>1$. This is illustrated in Fig. 1 where the solid
curve presents RHS, and the dashed straight lines present LHS
of Eq. \re{eq40} for $N_\gamma=0.1;0.3; 0.7;1.0;1.1$ from
the bottom  to the top, respectively.
It is seen that when
$N_\gamma<1$, there are two different solutions (denoted as crosses in Fig.1 )
 which overlap at
$N_\gamma=1$ and $Z=\pi/144=0.0218$, and then disappear.
This is one
of our main results confirming that there is a critical value of
$\gamma$, or more exactly critical value of $N_0\gamma/N$ which
controls the stability of the uniform Bose condensate at $T=0$.
When
$N_\gamma=432n_0\gamma/\pi$ exceeds unity, $(N_\gamma>1)$, $X_1$
and hence the self-energy becomes complex, and the BEC will be
unstable.

Differentiating Eq. \re{eq40} by $N_\gamma$ and solving with
respect to $dZ/dN_\gamma$, one obtains:
\be Z'\equiv
\dsfrac{dZ}{dN_\gamma}
=\dsfrac{\pi^{3/2}}{432(\sqrt{\pi}-12\sqrt{Z}) },
\lab{eq42} \ee
which is singular at $Z=\pi/144$, i.e., at
$N_\gamma=1$.
Thus, at the critical point, $N_\gamma=1$,
 \be
\ds{\lim_{N_\gamma\rightarrow 1} \dsfrac{\partial X_1}{\partial
n_0}}=\infty,
\lab{eq43}\ee
and hence, at this point the chemical potential of the condensate
$\mu_0$ in Eq.  \re{f42}, which is responsible
for the thermodynamical stability of the system,  has a singularity.

For $N_\gamma\leq 1$, the solutions are given as
\footnote{See e.g.
http://www.1728.com/cubic2.htm, about solving cubic equations for
the case when  Cardano's formula doesn't work.}
 \begin{eqnarray}
Z_1&=&\dsfrac{\pi}{576}[ 2c_1\cos(c_2)+3 ]
\nonumber \\
& \approx &
\dsfrac{\pi}{64}-\dsfrac{\pi}{216}N_\gamma+O(N_{\gamma}^{2}),
\lab{eq44}
\\
 Z_2&=&\dsfrac{\pi}{576}[-
c_1\cos(c_2)+\sqrt{3}c_1\sin(c_2)+ 3 ]
\\
 & \approx & \dsfrac{\pi
N_\gamma}{432}+\dsfrac{\sqrt{3}\pi N_\gamma^{3/2} }{1944}
+\dsfrac{\pi N_\gamma^{2} }{1944} +O(N_{\gamma}^{5/2}),
\lab{eq45}
\end{eqnarray}
where $c_1=\sqrt{9-8N_\gamma}$, $c_2=\arccos\{[27-36N_\gamma+8N_{\gamma}^2]/c_{1}^{3} \}/3$.

It is understood that only the second solution, $Z_2$, is a physical
one, since for the case of  $Z=Z_1$ the self-energy $X_1$ is
irregular at $\gamma\rightarrow 0$.
Moreover, only  $Z=Z_2$
corresponds to the minimum of the  thermodynamic potential,
$\ds{(\partial^2 {\Omega}/\partial^2 X_1   )|_{z=z_2}}>0.$
Thus, we  conclude,
$\xbar=2g\rho Z/\gamma$ with  $Z=Z_2$.
In
particular, taking into account the first term in the expansion
of $Z_2$ in Eq. \re{eq45}, one obtains $\xbar\approx 2g\rho
n_0=\xbar^{1L}$ as expected.

\begin{figure}
\resizebox{!}{0.25\textheight}{\includegraphics{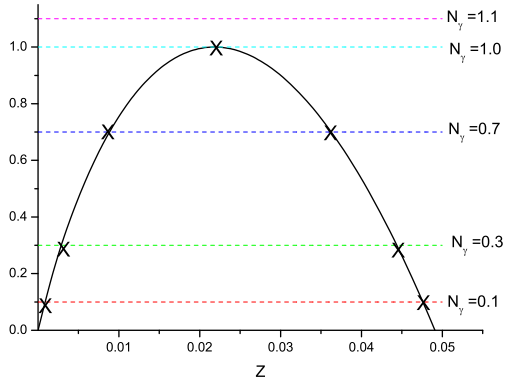}}
\caption{\label{fig:wide}
Graphical solution of the gap equation \re{eq40}.
     The solid curve represents RHS, and the dashed straight lines represent LHS
of the equation  for $N_\gamma=0.1;0.3; 0.7;1.0;1.1$ from
the bottom  to the top, respectively.
}
\end{figure}

\section{Results and discussions}

{\bf Expansion for small $\gamma$}.
The starting point of our
numerical calculations is the Eq. \re{eq39}, which can
be rewritten as
\be
1-{\nbar}_0-\dsfrac{8Z^{3/2}(\nbar_0)}{3\gamma\sqrt{\pi}}=0.\lab{rd1}\ee
Before
analyzing this  nonlinear equation we note that
the majority of experiments
with ultracold trapped gases deal with weakly
interacting atoms,
so that  $\gamma$ is very  small, i.e.  $\gamma\sim 10^{-9}\hdots 10^{-4}$.
Thus, to obtain a low- density expansion of physical quantities, one
may search for the solutions of Eq. \re{rd1} in  power series of
$\sqrt{\gamma}$ to get
\begin{eqnarray}
\nbar_1&=&1-\nbar_0
\nonumber \\
&=&\frac{8}{3}{\left(\frac{\gamma}{\pi}\right)}^{1/2}
+\frac{64}{3}\frac{\gamma}{\pi}+O(\gamma^{3/2}).
\lab{rd2}
\end{eqnarray}
Clearly, the first term corresponds to the  Bogoliubov approximation,
while the others may be considered as quantum
corrections to this approximation. Note also that the above expansion
Eq. \re{rd2} is exactly the same as the one obtained in the modified
Hartree-Fock Bogoliubov (HFB) approximation  \ci{yukalovkl}.
Now, using Eqs. \re{eq37}, \re{f42}, \re{davlen} , \re{f45}, \re{f49}, \re{f50} and \re{rd2} we obtain the following low density expansions  for the energy density, chemical potentials, self
energies, the sound velocity and the pressure:
\be
\ba
{\cale}\approx
g\rho^2\{1+\dsfrac{128\sqrt{\gamma}}{\sqrt{\pi}}+\dsfrac{128\gamma}{9\pi}
 \}
 , \quad
 {\mu_0}\approx
g\rho\{1+\dsfrac{32\sqrt{\gamma}}{3\sqrt{\pi}}+\dsfrac{224\gamma}{3\pi}
 \}.
 \lab{rd3}\ea \ee
\be {\mu_1}\approx
g\rho\{1-\dsfrac{16\sqrt{\gamma}}{3\sqrt{\pi}}-\dsfrac{128\gamma}{3\pi}
 \}  , \quad
 {\Sigma_{12}}\approx
g\rho\{1+\dsfrac{16\sqrt{\gamma}}{3\sqrt{\pi}}+\dsfrac{128\gamma}{3\pi}
 \}.    \lab{rd4}\ee

\be {c^2}\approx \dsfrac{g\rho}{m}
\{1+\dsfrac{16\sqrt{\gamma}}{3\sqrt{\pi}}+\dsfrac{128\gamma}{3\pi}
 \}, \quad
 {P}\approx \dsfrac{g\rho^2}{2}
\{1+\dsfrac{64\sqrt{\gamma}}{5\sqrt{\pi}}+\dsfrac{64\gamma}{3\pi}
 \}.  \lab{rd5}\ee
which are in good agreement with BPA \ci{dickbook}.

{\bf Critical density and exact solutions. }
In order to  discuss exact solutions of the  equation,  \re{rd1},
we first establish the
boundary for $\gamma$ which is related to the critical value of
$N_\gamma$ found in the previous section.
This may
be evaluated directly by substituting $N_\gamma=1$, $Z=\pi/144$
into Eq. \re{rd1}, which immediately gives
$\gamma_{cr}=5\pi/1296\approx 0.012120$. It is interesting to
observe that when $\gamma$ approaches  this critical value, the
condensed fraction remains still large,
$\ds{\lim_{\gamma\rightarrow\gamma_{cr}}{\nbar_0}}(\gamma)
=\pi/432\gamma_{cr}=3/5=0.6$
 but the condensate as a whole become unstable.

Fig. 2 presents the condensate fraction $\nbar_0(\gamma)$ in the Gaussian (solid line),
the one-loop (dotted line) and Bogoliubov-Popov approximations. It is seen that due
to the quantum fluctuations the condensed fraction decreases
faster (with increasing $\gamma$) in the Gaussian approximation
than in  BPA.
\begin{figure}
\resizebox{!}{0.25\textheight}{\includegraphics{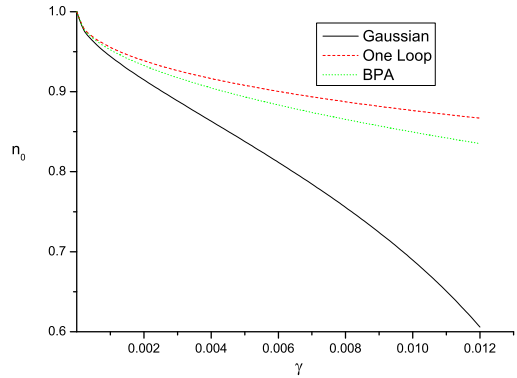}}
\caption{\label{fig:wide}
Condensate fraction $n_0=n_0(\gamma)$ as a function of $\gamma=\rho a^3$.
Solid, dashed and dot-dashed curves correspond to the  Gaussian,
the one-loop and Bogoliubov-Popov approximations.
}
\end{figure}



The chemical potential $\mu=\mu_0 n_0+\mu_1 n_1$ is presented
in Fig. 3.
One may observe that, in the Gaussian approximation it varies slowly
with increasing $\gamma$ , almost coinciding with that for the BPA.
\begin{figure}
\resizebox{!}{0.2\textheight}{\includegraphics{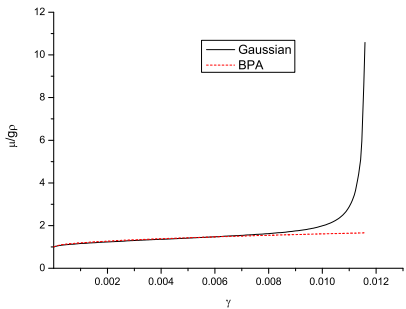}}
\caption{\label{fig:wide}
  The chemical potential in the Gaussian (solid line)
 and Bogoliubov-Popov approximations.
}
\end{figure}
However, when $\gamma$ approaches the critical value $\gamma_{crit}=0.012$ it starts to increase very fast since in this region, when $N_\gamma\rightarrow 1$,
$X_1'$ in Eq. \re{f42} becomes very large and so does $\mu_0$. Bearing in mind that
the chemical potential is the energy needed to add (or extract)
one more particle to (or from) the system, one may interpret this effect
as a particle number saturation of the condensed particles.
In other words, when $\gamma$ (or more exactly $N_\gamma$)
reaches the critical value, the number
of condensed atoms $N_0$ cannot be further increased, since it will lead
to a dynamical instability of the BEC.
The pressure defined by eq. \re{davlen} is positive
in the region $\gamma<\gamma_{crit}$, but  near the critical point $\gamma\sim\gamma_{crit}$ it becomes negative and small, as expected.
Note that at this point the energy density, the self-energies and sound velocity remain
finite, since corresponding expressions Eqs. \re{f45}-\re{f50}
 do not include $X'_1$ explicitly.

Now we consider possible origin of the instability found above. It is well known
that \ci{huangbook} the BEC is an effect of the  exchange coupling, which leads to an effective attraction between atoms forcing them to accumulate in a single state. However,
when the density (or scattering length) reaches a critical value this effective
attraction makes the condensate collapse.

Another possible reason is a three body recombination of condensed atoms.
Although  there is no explicit 3 body interaction
in our starting Lagrangian,
it was shown that  \ci{fedichev},
at $T\rightarrow 0$,  the repulsive  two body interaction leads to a three body
recombination with the rate constant $\alpha_{rec}\propto a^4$ and,
hence, the three body recombination  becomes very significant
for a large scattering length i.e. large $\gamma$.
This seems to be  one
of the main reasons for the fact that a stable  condensate with large
gas parameter is inaccessible experimentally. When $\gamma$
exceeds the critical value, the atoms start to combine into molecules and
 the condensate may undergo phase transition into a solid or a liquid state.

 \section{Summary}

 In conclusion, we have developed a new Bosonic self-consistent
variational perturbation
 theory, which can be made the starting point for
   systematic expansion procedure \ci{chikustev}.
We have shown that taking   into account two normalization conditions
at the same time  solves the old outstanding problem    of Bose systems
making variational perturbation theory both conserving and gapless.

   Studying the properties of a system of uniform Bose gas at zero
   temperature with repulsive interaction both analytically and numerically,
   we have found that in this system there is a dynamical parameter
    $N_\gamma\propto n_0\rho a^3$ which controls the stability of the Bose condensate.
    When  this parameter remains smaller than the critical value the phonon spectrum
    is purely real and the excitations have  infinite lifetimes. On the contrary, when
    $N_\gamma$ exceeds the critical value the condensate becomes unstable,
in      similar fashion to the BEC with an attractive interaction.
Note that this phenomena
     cannot be obtained in ordinary perturbative framework.

It would be quite interesting to study the dependence of critical $N_\gamma$ on
temperature.  It was discovered long ago by Bethe \ci{weiner} that the inelastic
cross section, which tends to destroy the condensate, varies as $1/k$,
(called as 1/velocity law), and hence the bad collisions can be surprisingly
large near zero temperature.  Thus, the temperature dependence of
the critical $N_\gamma$ seems not to be trivial. This work is on progress.

\section*{Acknowledgments}
  A.R.
appreciates  the Yonsei  University  for hospitality
 during his stay, where   the main part of
this work was  performed. We are indebted to V. Yukalov for several
constructive remarks and highly useful advice.
 This work was supported by the second phase of the Brain Korea21
  Project. C.K.K. acknowledges the support from the Korea Science
  and Engineering Foundation (R01-2006-000-10083-0).
\bb{99}
\bi{obzorkettel} W. Ketterle,  \rmp 
{\bf 74}, 1131 (2002).
\bi{nozarezkukkitob} P. Nozie'res, {\it Bose-Einstein Condensation},
    ed. by A. Griffin, D. W. Snoke and S. Stringari (Cambridge Univ., New York, 1995).
\bi{donleynature} E. A. Donley {\it et al.}, Nature, {\bf 412}, 295 (2001).
\bi{blochcondmat} I. Bloch, J. Dalibard and W. Zwerger,  e-print
    arXiv:cond-mat/0704.3011  (2007).
\bi{cornish} S. L. Cornish {\it et. al}. \prl 
 {\bf 85}, 1795 (2000).
\bi{Yin} L. Yin, e-print arXiv:cond-mat/0710.5318 (2007).
\bi{pitbook} L. Pitaevskii and S. Stringari, {\it Bose-Einstein Condensation}
	 (Oxford Univ., New York, 2003).
\bi{gauss} V. I. Yukalov, Moscow Univ. Phys. Bull. {\bf31},10 (1976);\\
A. Okopinska \prd 
{\bf 35}, 1835, (1987); \\
 P. M. Stevenson \prd
 {\bf 32}, 1389 (1985); \\
  A. Rakhimov and J. H. Yee, Int. J. Mod. Phys. A {\bf 19}, 1589  (2004);\\
 C. K. Kim, A. Rakhimov, and J. H. Yee,  Euro. Phys. J.  B {\bf 39}, 301 (2004);
\\
       C. K. Kim, A. Rakhimov, and J. H. Yee,   \prb 
       {\bf 71},  024518 (2005).
 \bibitem{amelino} G. Amelino-Camelia and So-Young Pi, \prd 
  {\bf 47}, 2356 (1993).
 \bi{blismastoof} M. Bijlsma and H. T. C. Stoof, \pra 
 {\bf  55} 498, (1997).
\bi{andersen} J. O. Andersen, \rmp 
{\bf 76}, 599 (2004).
\bi{yukalov} V. I. Yukalov, Phys. Rev. {\bf E72}, 066119, (2005).
\bi{yukalovkl}
V. I. Yukalov and H. Kleinert, \pra 
{\bf 73}, 063612, (2006).
\bi{nagaosa} N. Nagaosa, {\it Quantum field theory in condensed matter physics}
	(Springer, 2000).
 \bi{haugset} T. Haugset, H. Haugerud and F. Ravndal,
	Ann. Phys. {\bf 27}, 266 (1998).
\bi{braatentrap} E. Braaten, A. Nieto, \prb 
 {\bf 56}, 14745 (1997).
\bi{klcondmat} H. Kleinert, S. Schmidt and A. Pelster,
	 e-print cond-mat/0308561 (2003).
 \bi{braatenepj} E. Braaten and A. Nieto, Euro. Phys. J. B {\bf 11}, 143 (1999).
\bi{dickbook} W. H. Dickhoff and D. Van Neck, {\it Many-Body Theory Exposed}
	(World Scientific, 2005)
  \bi{huangbook} K. Huang, {\it Statistical Mechanics}, 2nd (Wiley, New York, 1987).
\bi{fedichev} P. O. Fedichev, M. W. Reynolds and G. V. Shlyapnikov
  \prl 
   {\bf 77}, 2921 (1996)
\bi{chikustev} S. Chiku and T. Hatsuda, Phys. Rev {\bf 58}, 076001 (1998);\\
I. Stancu and P. M. Stevenson, Phys. Rev. {\bf42}, 2710 (1990).
\bi{weiner} J. Weiner, V. S. Bagnato, S. Zillio and O. S. Julienne,
	\rmp 
 {\bf71}, 1 (1999)
\eb
\edc